\documentstyle[aps,preprint]{revtex}
\begin{document}
\draft
\title{
       Charge Transport in  Voltage-Biased Superconducting 
       Single-Electron Transistors       }
 \author{Jens Siewert and Gerd Sch\"on}
\address{
 Institut f\"ur Theoretische
 Festk\"orperphysik, Universit\"at Karlsruhe, \\
 76128 Karlsruhe,  FRG\\
        }
\date{\today}
\maketitle
\begin{abstract}
    Charge is transported through superconducting SSS
    single-electron transistors at finite bias voltages
    by a
    combination of coherent Cooper-pair tunneling 
    and quasiparticle tunneling. At low
    transport voltages the effect of an ``odd''
    quasiparticle in the island
    leads to a $2e$-periodic dependence
    of the current on the gate charge.
    We evaluate the $I-V$ characteristic in the framework
    of a model which accounts for  these effects
    as well as for the influence of the electromagnetic
    environment. 
    The good agreement between our model calculation
    and experimental results demonstrates the importance
    of coherent Cooper-pair tunneling and parity effects.
\end{abstract}
The single-electron tunneling (SET) transistor has proven
an ideal system to display the concepts of ``single 
electronics''.
In this device an island is coupled via tunnel junctions
to the leads. The island potential can be modulated by 
a capacitively coupled gate voltage $V_g$. In transistors
with normal-conducting islands and leads the current depends
$e$-periodically on the gate charge $Q_0 = C_g V_g$.
Recently much attention has been devoted
to single-electron transistors with superconducting islands 
and normal leads (NSN) and entirely superconducting systems (SSS).
In a superconducting island, where Cooper pairs form
the condensate, the addition of one extra
electron - the ``odd'' one - costs the gap energy $\Delta$ 
\cite{Av+NazPRL92}. Hence the physical properties
of the system depend on the
parity of the charge number in the island, and the
 $I-V$ characteristics is expected to be  $2e$-periodic 
in the gate charge. The first clear signature of 
$2e$-periodicity  
was observed by Tuominen {\em et al.\ } \cite{Tuominen92}. 
However, no satisfactory explanation of these experimental
results has been provided until now. In the meantime a number 
of  experiments
both in NSN transistors \cite{Eiles,Hergshield} and in
SSS transistors \cite{SaclaySSS,Amar,Eiles2,JoyezTh,Delft}
have demonstrated a rich variety of phenomena and show 
good agreement with theoretical results.
It is now also well understood \cite{Hergshield} that the 
difficulties in observing the $2e$-periodicity arise from 
the extreme sensitivity of the even-odd difference to effects
of the electromagnetic environment which create non-equilibrium
quasiparticles. 

Parallel to the experiments the theoretical description of systems
exhibiting parity effects made rapid progress. In Ref.\ \cite{Tuominen92}
the authors present an equilibrium model which accounts for the
temperature dependence of the even-odd  asymmetry.
A kinetic model was developed in Ref.\ \cite{Gerd+Andrei}
 to describe transport. In this framework
the $I-V$ characteristics of NSN transistors
can be derived \cite{Hergshield,GJA}.
For SSS transistors, there exists a well-developed
theory  for the current-biased system
\cite{GlazmanSSS,SaclaySSS,Christoph}. The voltage-biased 
system  has been considered in the absence of parity effects
\cite{AMvdB}, while in Ref.\ \cite{JoyezTh} only resonant Cooper-pair
tunneling has been studied. In this paper we
investigate a model which includes  Cooper-pair 
and single-particle tunneling as well as parity effects.
First we study the $I-V$ characteristic of an SSS transistor
in the absence of  external impedances.
In the transport voltage range 
$eV \raisebox{-.3ex}{$\stackrel{<}{\sim}$} E_{C}$
even-odd effects are observed ($E_{C}$ denotes the scale of the charging
energy, see below). We then discuss examples for the relevant transport
processes. In order to compare with the experiment,
we account for the influence of the electromagnetic 
environment.  This rather complex model
explains the experimental results of Ref.\ \cite{Tuominen92}.

We consider a SET transistor (see Fig.\ 1) with superconducting 
electrodes and island (with energy gaps $\Delta$)
below the crossover temperature $T_{cr}=\Delta/k_{B}\ln{N_{eff}}$
where parity effects can be observed. 
Here 
$N_{eff}=2N_{I}(0)\sqrt{2\pi\Delta k_{B}T}$ is the effective number
of states available for the odd quasiparticle \cite{Tuominen92} and
$N_{I}(0)$ is the density of states (per spin) in the island.
For  the moment we ignore the effect of the external impedance.

The system can be described by a sum of Hamiltonians for the left
and right electrode and the island, the charging energy and the 
tunneling Hamiltonian 
$H=H_{L}+H_{R}+H_{I}+H_{ch}+H_{T}$.
We will treat the Josephson tunneling non-perturbatively.
Hence we start from the model Hamiltonian \cite{AMvdB}
\begin{equation} 
    \begin{array}{cl}
      H_0 = & \sum\limits_{Q,\bar{Q}}\left(\left[
       		     {\displaystyle \frac{(Q+Q_0)^2}{2C}-
      \frac{1}{2}\bar{Q}V} \right] |Q,\bar{Q} \rangle 
                     \langle Q,\bar{Q}| 
      \right.    \\[5mm]
      & \left.    {\displaystyle
      - \frac{E_J}{2}\sum_{\pm} \sum_{\pm}
      |Q\pm 2e,\bar{Q}\pm 2e \rangle \langle Q,\bar{Q}| } \right)
    \end{array}
\label{hnull}
\end{equation}
and we will account  for the quasiparticle tunneling in perturbation
theory.
Here $Q\equiv (n_l - n_r)e$ denotes the island charge,
and we defined a total charge  which has passed through 
the  system $\bar{Q}\equiv (n_l + n_r)e$ 
(with respect to a reference state). 
Further  $C = C_l + C_r + C_g$ is the total island capacity 
and $Q_{0}=C_{g}V_{g}$  the offset charge (here we 
assume $C_{l}=C_{r}$). 
The first two terms describe the charging energy and the
energy gain in tunneling due to the transport voltage.
They are diagonal in the basis of charge states $|Q,\bar{Q}\rangle$.
The typical scale of the charging energy
is $E_{C}=e^{2}/2C$. 
The last term describing Cooper-pair tunneling
with coupling energy $E_{J}$ is off-diagonal 
and, thus,  mixes the charge states $|Q,\bar{Q}\rangle $.
Therefore, the eigenstates of $H_{0}$ are linear combinations
of these states
\begin{equation}
 |\Psi_k \rangle = \sum\limits_{n,m} a^k_{n,m} |ne,me \rangle \ \ . 
\label{psih0}
\end{equation}

A dc-current requires a dissipation mechanism. 
In the case of zero external impedance the quasiparticle tunneling
can cause transitions between different eigenstates $|\Psi_k\rangle$.
It is accounted for by 
\begin{equation}
\begin{array}{ccl}
        H_{T}^{qp} & = & {\displaystyle 
                  \sum\limits_{k \in I, 
                         p \in L}} 
	          T^{(l)}_{pk} |Q+e,\bar{Q}+e\rangle \langle Q,\bar{Q}| 
                  c_k^{\dagger} c_p + \mbox{h.c.}     \\[3mm]
                  && \hspace*{-6.5mm}
                  + {\displaystyle \sum\limits_{k \in
                  I, p'\in R}} 
                   T^{(r)}_{kp'} |Q-e,\bar{Q}+e\rangle \langle
	          Q,\bar{Q}|
                  c_{p'}^{\dagger} c_k + \mbox{h.c.} 
\end{array}
\label{Hqp}
\end{equation}
The tunneling matrix elements in the left/right junction $T^{(l/r)}_{pk}$
are related to the conductance of the junctions by 
%
$
1/R_{l/r} =  4\pi e^{2}/\hbar\ N_{L/R}(0)N_{I}(0)
                        |T^{(l/r)}|^{2} 
$,
%
where we approximated $T^{(l/r)}_{pk}\approx T^{(l/r)}$.
If the  junction resistances are large compared to the
quantum resistance $R_{l/r}>R_{K}=h/e^{2}$
the transition rates can be calculated by the golden rule
\begin{equation}
  \begin{array}{cl}
   \Gamma_{i \rightarrow f} = & {\displaystyle \frac{1}{e}
                \sum\limits_{j,Q,\bar{Q}}  \left(
   \frac{I_{qp}^{(j)}(\varepsilon_{if})}{1 - \exp(-\varepsilon_{if}/k_{B}T)}
                   +  e\gamma _{esc}  \right)}
     \ \ \cdot  \\[3mm] &     
       \hspace*{-10mm}\cdot   {\displaystyle \sum\limits_{j=l:\ \pm\pm}
                            \ \sum\limits_{j=r:\ \pm\mp}}
   |\langle \Psi_f | Q\pm e,\bar{Q}\pm e \rangle \langle 
   		     Q,\bar{Q}       	     
   | \Psi_i \rangle |^2 \ .
  \end{array}
\label{qprate}
\end{equation}
Here $I_{qp}^{(j)}$ is the well-known $I-V$ characteristic for
quasiparticle tunneling \cite{Mike} between superconductors 
and $\varepsilon_{if}$
is the energy difference between initial and final state.
In order to allow for the parity effect, we have to 
include 
the escape rate $\gamma_{esc}$ of an odd quasiparticle in the
island \cite{Gerd+Andrei} 
\begin{equation}
   \begin{array}{cl}
\gamma_{esc} \simeq  & \\[5mm]
		      & \hspace{-1.1cm} \left\{
                               \begin{array}{ll} 
                       {\frac{1}{2 e^2 R_{l/r} N_I(0)}
                        \frac{\varepsilon_{if} +\Delta}
                       {\sqrt{(\varepsilon_{if} +\Delta )^{2}-\Delta^{2}}}
		       } \theta (\varepsilon_{if}) \  & 
                                           \mbox{if}\   Q\ \mbox{odd} \\
                          0                                  &
                                           \mbox{if}\   Q\ \mbox{even}       
                               \end{array}  
                        \right. \ .  
   \end{array}
\label{gesc}
\end{equation}

In order to determine the dc-current, we follow the procedure
described in Ref.\ \cite{AMvdB}: First, we determine the
eigenstates of $H_{0}$  either in perturbation
theory (as we shall discuss below) or numerically 
taking into account a sufficient number of charge states.
This procedure converges for not too large Josephson coupling energies
$E_{J} < E_{C}$. 
Given the eigenstates $|\Psi_{k}\rangle$ we calculate 
the rates in Eq.\ (\ref{qprate}), which then enter
a master equation 
%
\mbox{$
\partial_{t} P_{k} = \sum_{n\neq k}
                             (P_{n}\Gamma_{n\rightarrow k}
                           -  P_{k}\Gamma_{k\rightarrow n})
$}
%
for the probabilities $P_{k}$ to find
the system in the $k$-th eigenstate. The stationary solution
$\partial_{t} P_{k} = 0$ is sufficient to evaluate the
dc-current
\begin{equation}
I =    \frac{e}{2}
       \sum\limits_{k, n\neq k} P_{k}\Gamma_{k\rightarrow n}
                       (\langle \Psi_{n} |\bar{Q}| \Psi_{n} \rangle
                     -  \langle \Psi_{k} |\bar{Q}| \Psi_{k} \rangle) \ \ .
\end{equation}
The results are shown in Fig.\ 2(a). 
We used the parameters
$\Delta = 1.3 E_{C}$, $E_{J}^{0}=0.17 E_{C}$, $R_{l/r}=R\approx R_{K}$,
$\gamma_{esc}=2.5\cdot 10^{-5} (RC)^{-1}$, which correspond to
those in Ref.\ \cite{Tuominen92}.
Here $E_{J}^{0}$ denotes the Ambegaokar-Baratoff expression 
for the Josephson energy. In Eq.\ (\ref{hnull}) the
generalized Josephson coupling energy in 
the presence of charging effects \cite{GAPRep} enters, which 
in the present case is larger than $E_{J}^{0}$ by roughly 20 \%.
We note that the $I-V$ characteristic is $2e$-periodic and
observe a rich structure deep in the subgap region. 
For transport voltages $eV \raisebox{-.3ex}
{$\stackrel{>}{\sim}$}
2.5 E_{C}$ the $2e$-periodic features disappear and the current 
becomes  $e$-periodic in $Q_{0}$ again.
This is not surprising since on a current scale 
$I\gg e\gamma_{esc}$ the  unpaired quasiparticle 
in the island looses its importance.

The basis of eigenstates of $H_{0}$ is ``appropriate'' for the problem.
On inspecting our numerical procedure we find that for
low transport voltages only a few
(two or three) states $|\Psi_{k}\rangle$ are noticeably populated.
Similar behavior is found in systems without
coherent tunneling like NNN or NSN transistors.
Therefore, we can calculate  transition rates and the current 
analytically if we know the eigenvalues of $H_{0}$ and the corresponding
eigenstates. To this end, we determine the coefficients $a^k_{n,m}$
in Eq.\ (\ref{psih0}) by using  perturbation theory in $E_{J}$.
Away from certain resonant situations, the $k$-th eigenstate
has only one coefficient $a_{q,p}^{k}$ of  order
unity, whereas all other coefficients are considerably smaller.
To fix ideas, let us consider the state $|\Psi_{0}\rangle$
 in the range of gate charges
$Q_{0}\in [0,e/2]$. In this eigenstate 
the most likely  charge state is 
$|Q=0e,\bar{Q}=0e\rangle$, i.e.\ $a_{0,0}^{0}\approx 1$. 
Due to coherent tunneling of one Cooper pair, 
there is a small amplitude $a_{\pm2,\pm2}^{0}\propto E_{J}/E_{C}$ 
to find the system
in the charge states $|Q=\pm 2e,\bar{Q}=\pm 2e\rangle$. Since also
several Cooper pairs can tunnel coherently, the system can be, e.g.,
in the charge state  $|2e,6e\rangle$ 
with a very small amplitude $a_{2,6}^{0}\propto (E_{J}/E_{C})^{3}$.

At resonance lines, however, it is possible that this
amplitude is much larger. Let us consider
the solid straight line in Fig. 2 (b), which is given by
\begin{equation}
    3eV = 4E_{C}(1-Q_{0}) 
 \ \ .
\label{line3}
\end{equation}
Along this line the charge states $|0e,0e\rangle$ and $|2e,6e\rangle$
have the same energy, i.e., three Cooper pairs tunnel resonantly
there. The resonance results in a drastically increased amplitude
$a^{0}_{2,6} \propto (E_{J}/E_{C})$.

A transition from $|\Psi_{0}\rangle$ to another eigenstate
can occur if it is energetically favorable and the matrix 
element of the final state with 
$|\Psi_{0}\rangle$ according to Eq.\ (\ref{qprate}) is nonzero.
On analyzing which transitions due to quasiparticle tunneling are 
energetically favorable, we find that a process
\[ |\Psi_{0}\rangle \approx |0e,0e\rangle 
         \longrightarrow |\Psi_{1}\rangle \approx |1e,7e\rangle 
              \hspace{1cm}    \mbox{(process a)} \]
is possible.
Out of resonance the rate of  process a) is of the order
$(E_{J}/E_{C})^{6}$. In a narrow strip (whose width is characterized by
$E_{J}$) around the resonance line Eq.\ (\ref{line3}), however, 
we find 
\begin{equation}
\Gamma_{a} \propto\left( \frac{E_J/2}{4E_C(1-Q_0)+eV} \right)^2  
           \propto \left(\frac{E_{J}}{E_{C}}\right)^{2}\ \ .
\end{equation}
The line in Fig. 2(b) corresponding to Eq.\ (\ref{line3}) marks 
the most significant resonance in the $I-V$ characteristic.
We are, thus, led to the conclusion that the dominant transport
process in the subgap region is tunneling of quasiparticles
accompanied by simultaneous tunneling of several
Cooper pairs. Due to this combination
enough energy is gained to overcome
the quasiparticle tunneling gap $2\Delta $. 
The importance of this type of transport mechanism was first
noted by Fulton {\em et al.\ }\cite{Fulton}.
Although the rates
for these processes in general are small, they are considerably
enhanced in situations, where resonant transfer of Cooper pairs
is possible.

So far we have studied the conditions for the system
to leave the initial state. However, a dc charge  
transport through the system requires cycles, after which the
island returns to a state equivalent to the initial one. 
The simplest
version is a two-step cycle of subsequent transitions 
of the same type in the left and right junction.
Such cycles dominate in NNN or NSN transistors
at low bias voltages. 
The cycles which lead to the pronounced features in Fig.\ 2 arise
due to two-step cycles as well, but the second step is different
from the first one.
The transition completing the cycle which starts with process a)
is
\[       |\Psi_{1}\rangle \approx
         |1e,7e\rangle \longrightarrow 
         |\Psi_{2}\rangle \approx |0e,12e\rangle 
                    \hspace{1cm}           \mbox{(process b)}\ \ , \]
i.e., quasiparticle transfer accompanied by only two Cooper pairs
tunneling.
The latter process is not in resonance and,
therefore, the rate is  $\Gamma_{b}\propto (E_{J}/E_{C})^{4}$. 
Whereas off-resonance the \mbox{process a)} is the bottleneck 
for the current,
at resonance the \mbox{process b)} has the smaller rate.
This explains that at the resonance the current
increases by roughly two orders of magnitude.

Another interesting feature in the $I-V$ characteristic
is the shoulder-like structure
between the high resonances for gate charges $Q_{0}\in [e/2,e]$. 
It  is directly
related to the escape rate $\gamma_{esc}$ of the odd particle.
The first step in the relevant cycle is a process similar to
\mbox{process b)}
\[          |\Psi_{0}\rangle \approx
            |0e,0e\rangle \longrightarrow 
            |\Psi_{3}\rangle \approx |1e,5e\rangle 
                     \]
with a rate  $\propto (E_{J}/E_{C})^{4}$, which is 
relatively large (as discussed before).
The current, however, is limited by the second step
\[        |\Psi_{3}\rangle \approx
          |1e,5e\rangle \longrightarrow 
          |\Psi_{4}\rangle \approx |0e,6e\rangle 
                    \ \ . \]
This is a pure quasiparticle transition without
Cooper pairs, which can occur because the escape rate $\gamma_{esc}$
has no gap. 
In Fig.\ 2(b) it is seen that the cycle sets in for
transport voltages $eV \geq 4E_{C}(Q_{0}-1/2)$ (the dashed
straight line).
This is exactly the condition for the odd quasiparticle
to gain energy on leaving the island.

So far we have considered the ideal case of a 
vanishing external impedance $Z$. In order to compare
our results with experimental data it is necessary to
account for the effect of the electromagnetic environment. 
An external impedance gives rise to 
incoherent Cooper-pair transitions.
For not too low transport voltages the rate of
these transitions is given by (see, e.g., 
Refs.\ \cite{AMvdB,Ing+Naz})
\begin{equation}
\Gamma_{i\rightarrow f}^{env} = \frac{1}{\hbar^{2}}
                     |\langle \Psi_{f}|\bar{Q}/2|\Psi_{i}\rangle |^{2} 
                           \frac{2\mbox{Re}Z(\omega )\ \varepsilon_{if}}
                           {1-\exp{(-\varepsilon_{if}/k_{B}T)}} \ \ .
\end{equation}
For appropriate parameters they lead to
a pronounced resonance structure in the $I-V$ characteristic
\cite{resonant}.

Furthermore, in an experiment the temperature of the environment
is not necessarily the same as the electron temperature
\cite{Mart+Nah,Hergshield}. 
High temperature fluctuations in the environment 
induce quasiparticle tunneling and, thus, can cause qualitative 
changes in the $I-V$ characteristic.
In order to take into account 
this effect,
we have to add to the single-electron tunneling rate
$\Gamma_{i\rightarrow f}$ the term
\begin{equation}
\Gamma_{i\rightarrow f}^{e,qp}\ =
	   \ \frac{1}{e^{2}R_{t}}\frac{2R_{e}}{R_{K}}k_{B}T_{e}
           \frac{2\Delta }{2\Delta -\varepsilon_{if}}
           e^{-(2\Delta - \varepsilon_{if})/k_{B}T_{e}}             
\label{highTqprate}
\end{equation}
for $\varepsilon _{if} < 2\Delta$.
The Ohmic resistance $R_{e}$ describes the environment,
which fluctuates at the high temperature
$T_{e}$\raisebox{-.3ex}{$\displaystyle\stackrel{>}{\sim}$}$E_{C}$. 
Since there is no detailed information about the electromagnetic 
environment in Ref.\ \cite{Tuominen92}, we  fix the
parameters by comparing with similar experiments. 
In Fig.\ 3 the $I(Q_{0})$ dependence for several transport voltages
is plotted. The calculation reproduces remarkably well both the general
shape of the experimental curves and the order of magnitude of
the current. We have cut the current resonances for
the bias voltages $eV=167\mu V,\ 200\mu V$, 
since the  maximum differential conductance, which can be observed in 
an experiment, is limited \cite{JoyezTh}.
In a model calculation like ours  we cannot expect perfect 
correspondence between theory and experiment.
The reason is that the system is  very sensitive
to even small amounts of quasiparticle tunneling
which can overcome a threshold for cycles in lower order of
$E_{J}/E_{C}$. (We have discussed a similar effect in connection with
the shoulder-like structure). Different relevant temperatures in 
the environment  (which is very likely)
can cause subtle changes in the $I-V$ characteristics.
In this case  Eq.\ (\ref{highTqprate}) is only a rough 
approximation.  Therefore, we have to adjust the parameter $R_{e}$   
for different bias voltages.
Finally we mention that one can speculate about 
processes of higher order in the quasiparticle
tunneling, e.g., co-tunneling or coherent two-electron tunneling
across one junction, 
accompanied by Cooper pairs. However, we come to the conclusion
that in the present case processes of higher order in the 
quasiparticle tunneling can be neglected.
There are two possibilities: i) The process creates excitations.
Then the rate has a gap $\geq 2\Delta$
and becomes important only for higher transport voltages,
where also first order quasiparticle tunneling is possible.
ii) The process does not create excitations.
But then  the phase space is reduced
by at least one  small factor  $1/(N_{I}(0)\Delta )$ \cite{Av+NazPRL92}. 

Summarizing, we can say that our non-perturbative model 
adequately describes charge transport in voltage biased
SET transistors under various conditions in a large range
of transport voltages.  Due to the parity effect the $I-V$ 
characteristics are $2e$-periodic in the gate charge for not 
too large transport  voltages. The dominant
transport mechanism is quasiparticle tunneling, accompanied
by coherent tunneling processes of several Cooper pairs.
An  external impedance at low temperatures causes incoherent  
Cooper-pair transitions, whereas high temperature fluctuations in
the environment ``poison'' the ideal structure, thus 
restoring $e$-periodicity of the current.
%
%
%
%

We acknowledge stimulating discussions
with R.\ Bauernschmitt and A.\ Rosch and thank
P.\ Joyez for a copy of his thesis. This work is
supported by ``Sonderforschungsbereich 195'' of
the Deutsche Forschungsgemeinschaft.
\begin{figure}
\caption{The SSS transistor. $n_{l/r}$ is the number
         of electrons which have tunneled through the left and
         the right junction.
         }
\end{figure}
\begin{figure}
\caption{a) $I-V$ characteristic of a SSS transistor for
          vanishing external impedance (parameters see
          text). 
          b) Contour plot of the same data.
          There are three dashed contour lines in the current range
          $I=0\ldots 10^{-5}\ e/(RC)$ and 20 lines for $I\leq 10^{-3}
          \ e/(RC)$. A pronounced resonance is found along
          the straight solid line 
          $3eV=4E_{C}(1-Q_0)$. The straight dashed line 
          $eV = 4E_{C}(Q_{0}-1/2)$ marks the edge of the
          shoulder-like structure.}
\end{figure}
\begin{figure}
%
%
\caption{The current $I(Q_{0})$ through an SSS SET transistor,
          including both resonant Cooper-pair transitions
          due to an external impedance $Z$ and high temperature
          fluctuations in the electromagnetic environment.
          Parameters are: $E_{C}=180\mu$eV, $\Delta=240\mu$eV,
          $E_{J}^{0}=30\mu$eV, $R_{l}+R_{r}=50$k$\Omega$,
          $\gamma_{esc}\simeq 2\cdot 10^{6}s^{-1}$,
          Re$Z(0)$=80$\Omega$, $R_{e}=0.8\Omega$ (for
          $V\leq 123\mu$V), $R_{e}=3\Omega$ (for $V\geq 150\mu$V),
          $T_{e}$=3.6K. We have assumed a slight asymmetry of
          the junction resistances $R_{l}:R_{r}=4:3$.
          }
\end{figure}

\begin{references}
\bibitem{resonant} 
    We note that in Ref.\ \cite{JoyezTh} resonant Cooper-pair 
    tunneling has been observed in a wide range of transport 
    voltages. The results 
    have been explained there within 
    a less complex model for resonant Cooper-pair tunneling.
    We can reproduce the experimental data of Ref.\ 
    \cite{JoyezTh}  in a range  $eV=0.2E_{C}\ldots 4E_{C}$
    (not shown here), although
    the Josephson coupling energy in the experiments is of the
    order of the charging energy.
\end{references}
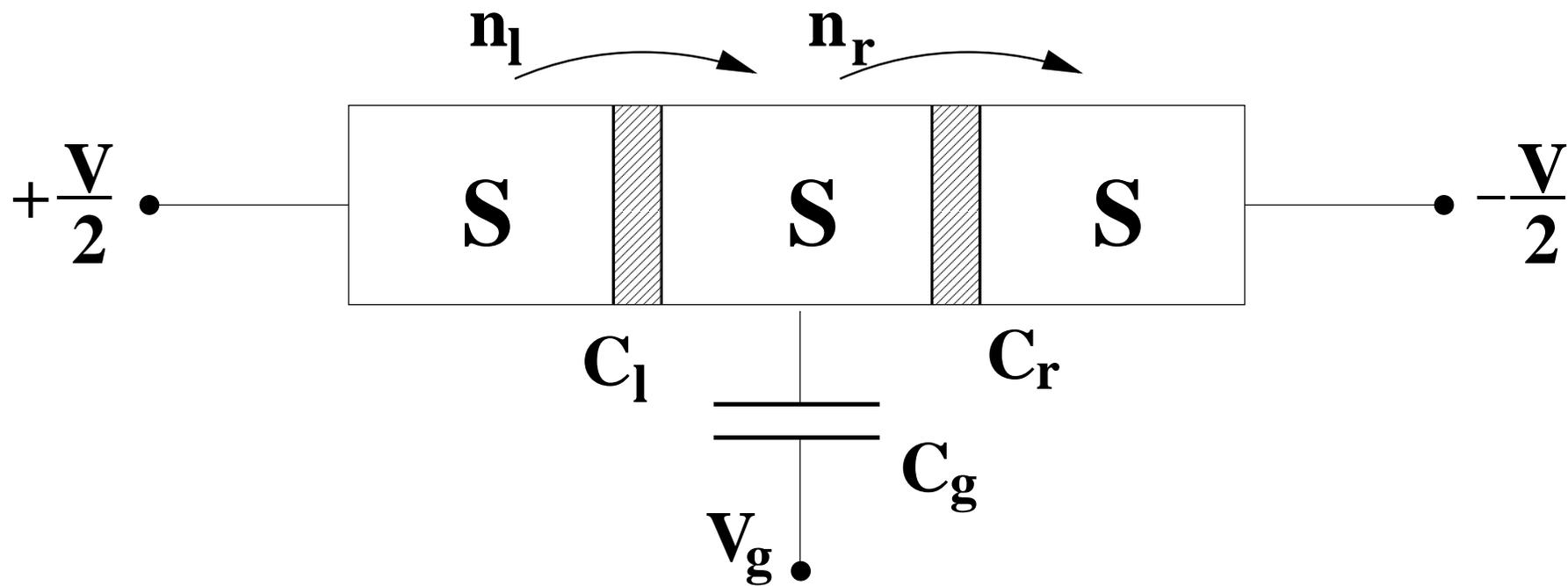
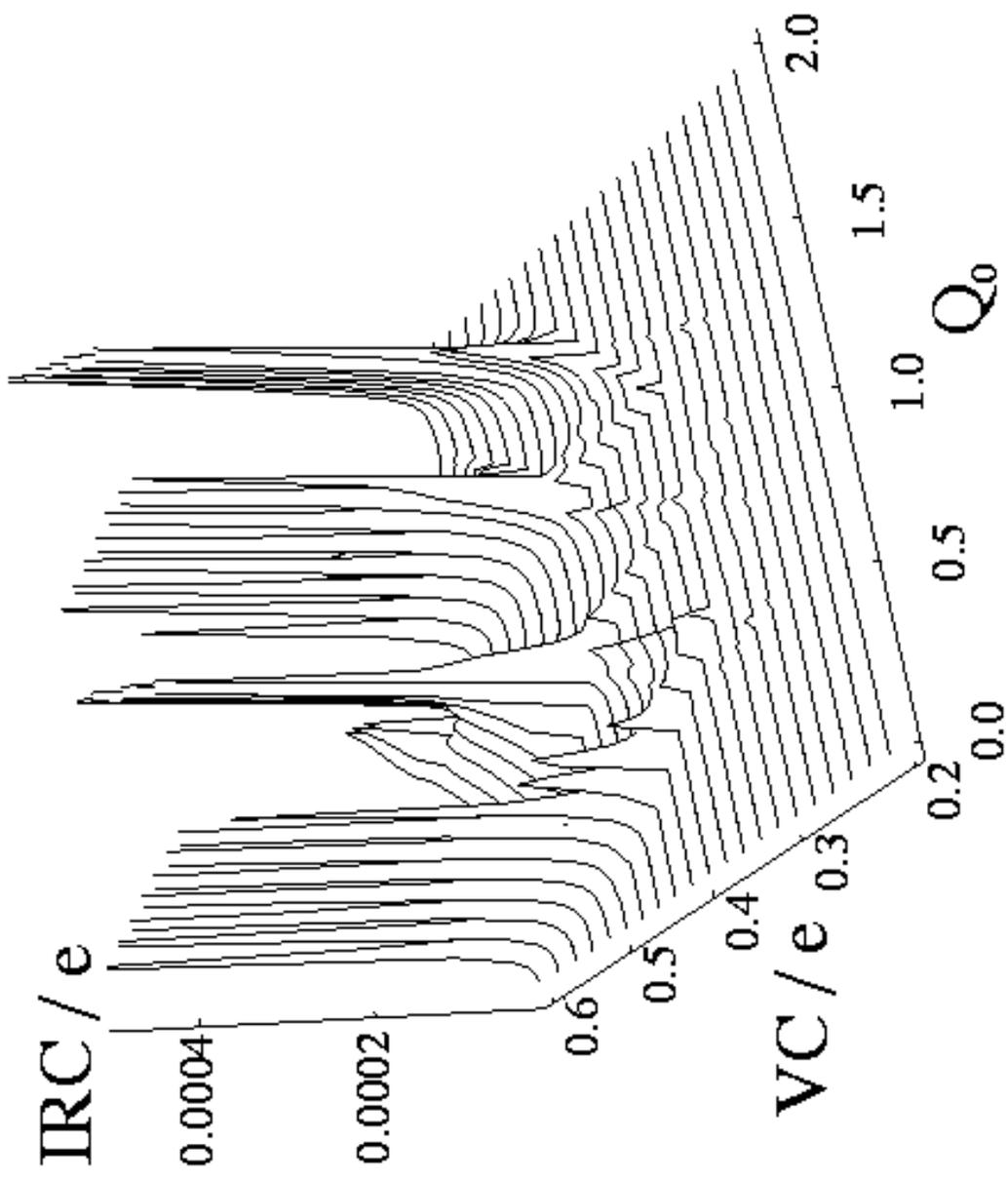
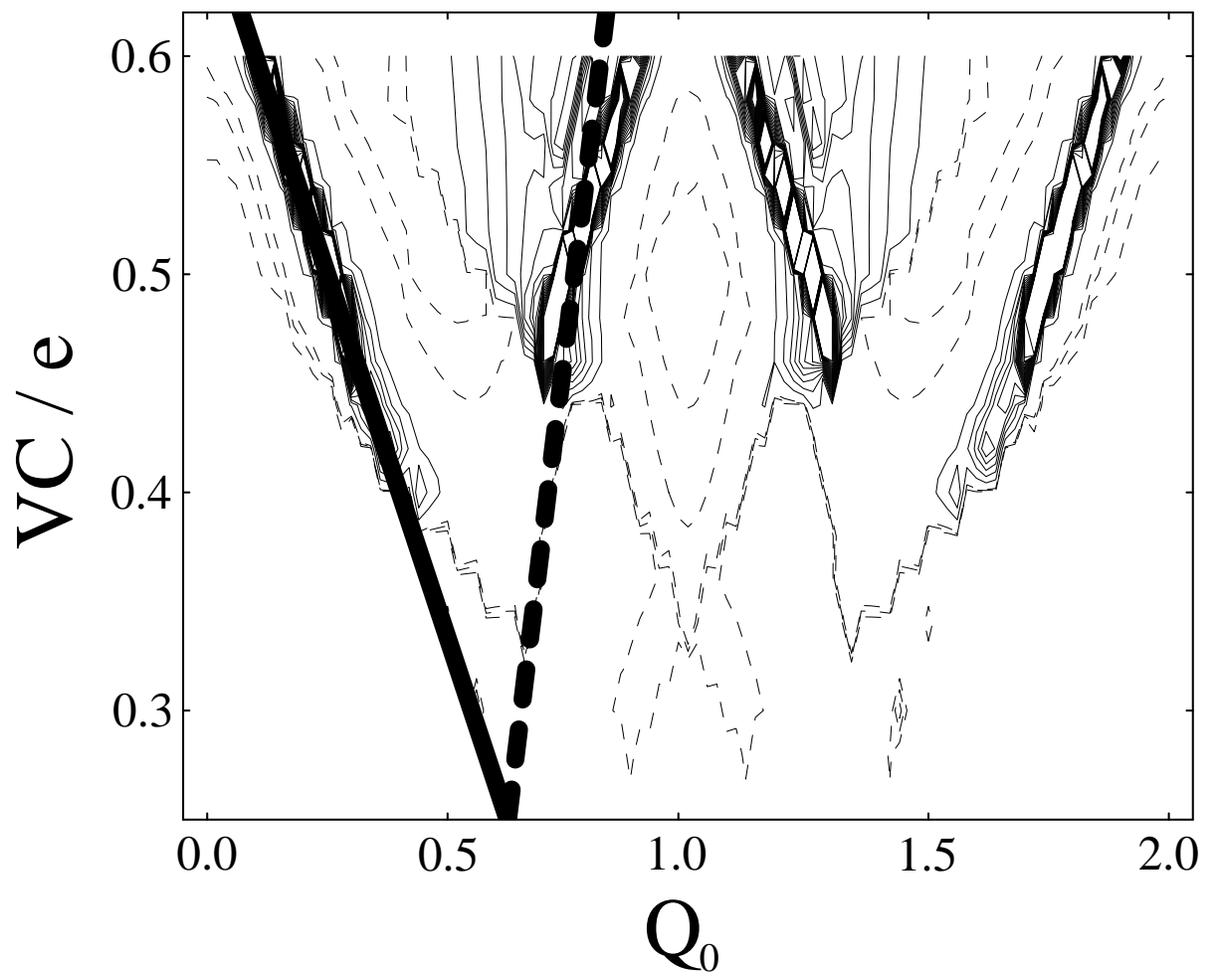
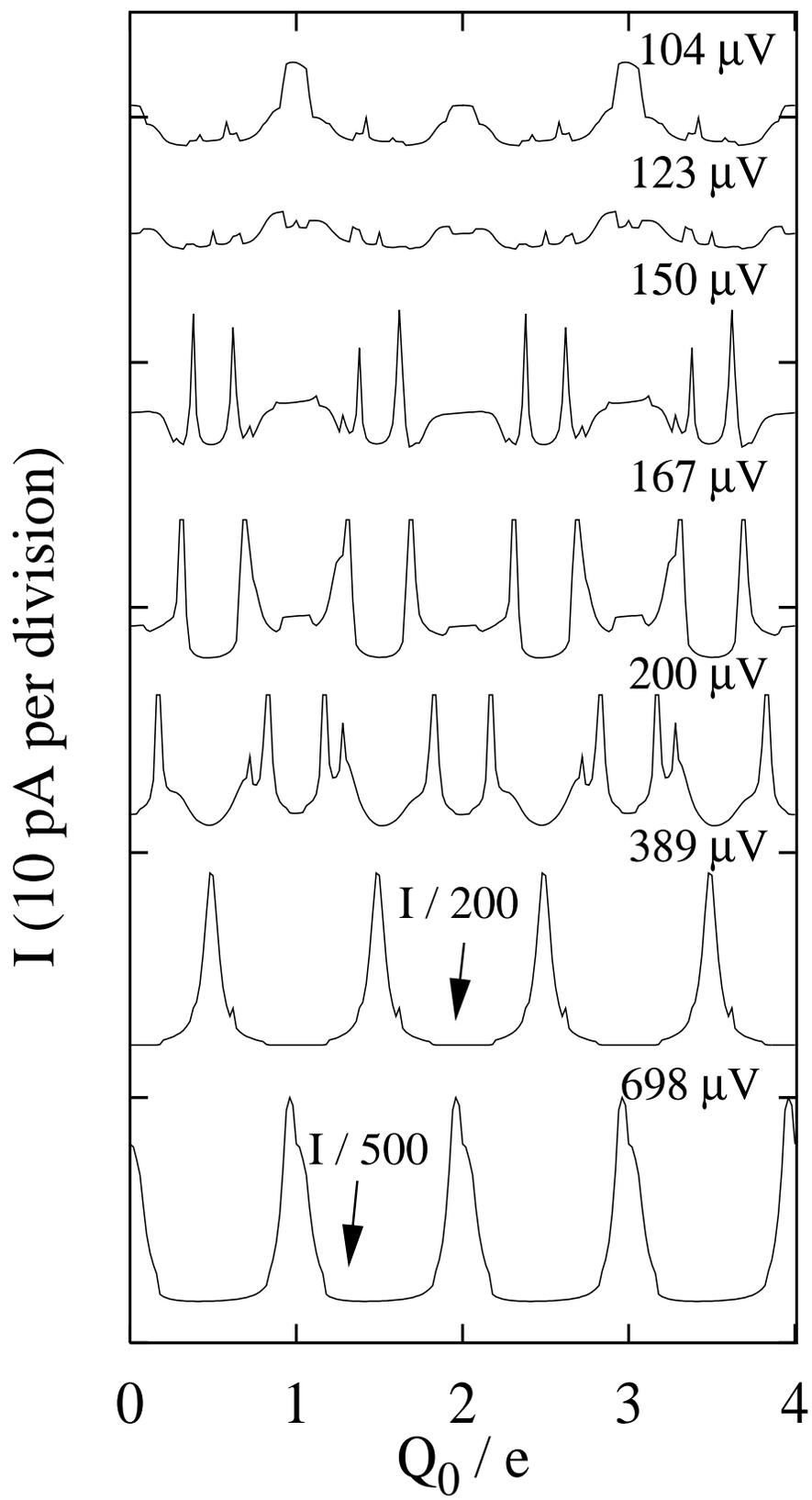
\end{document}